\documentclass{emulateapj}
\providecommand{\mu}{$\Delta m_{15}(U)$}
\begin{document}
%%%%%%%%%%%%%%%%%%%%%%%%%%%%%%%%%%%%%%%%%%%%%%%%%%%%%%%%%%%%%%%%%%%%%%%%%%%%%%%%%%%
%%%%%%%%%% Title & Author %%%%%%%%%%%%%%%%%%%%%%%%%%%%%%%%%%%%%%%%%%%%%%%%%%%%%%%
\title{Ultraviolet Light Curves of Supernovae with {\sl Swift} UVOT}
%%%%%%%%%%%%%%%%%%%%%%%%%%%%%%%%%%%%%%%%%%%%%%%%%%%%%%%%%%%%%%%%%%%%%%%%%%%%%%%%%
%%%%%%%%%%%%%%%%%%%%%%%%%%%%%%%%%%%%%%%%%%%%%%%%%%%%%%%%%%%%%%%%%%%%%%%%%%%%%%%%%%
\author{Peter~J.~Brown\altaffilmark{1}, 
Stephen T. Holland\altaffilmark{2,3,4}, 
Stefan Immler\altaffilmark{2,4}, 
Peter Milne\altaffilmark{5},
Peter~W. A.~Roming\altaffilmark{1}, \\
Neil Gehrels\altaffilmark{2}, 
John Nousek\altaffilmark{1},
Nino Panagia\altaffilmark{6,7,8},
Martin Still\altaffilmark{9},
\& Daniel Vanden Berk\altaffilmark{1}
}

\altaffiltext{1}{Pennsylvania State University,
               Department of Astronomy \& Astrophysics,
              University Park, PA 16802, USA}                             
\altaffiltext{2}{Astrophysics Science Division, Code 660.1,
      NASA/Goddard Space Flight Center, Greenbelt, MD 20770, USA }
\altaffiltext{3}{Universities Space Research Association,
      10227 Wincopin Circle, Suite 221,
      Columbia, MD 21044, USA}
\altaffiltext{4}{Centre for Research and Exploration in Space Science and Technology
      Code 660.8, NASA/GSFC, Greenbelt, MD 20770, USA}  
%\altaffiltext{7}{INAF, Observatorio Astronomico di Trieste}
%\altaffiltext{5}{Max-Planck-Institut f\"ur extraterrestrische Physik}
\altaffiltext{5}{Steward Observatory, University of Arizona, 
933 North Cherry Avenue, Tucson, AZ 85721}              
\altaffiltext{6}{Space Telescope Science Institute, 3700 San Martin Drive,
Baltimore, MD 21218, USA}
\altaffiltext{7}{INAF-Osservatorio
Astrofisico di Catania, Via S. Sofia 78, I-95128 Catania, Italy}
\altaffiltext{8}{Supernova Ltd., OYV \#131, Northsound Road, 
Virgin Gorda, British Virgin Islands}
\altaffiltext{9}{Mullard Space Science Laboratory,
		 Department of Space and Climate Physics,
		 University College London, Holmbury St Mary,
		 Dorking, Surrey, RH5 6NT,
		 UK}

%%%%%%%%%% Abstract %%%%%%%%%%%%%%%%%%%%%%%%%%%%%%%%%%%%%%%%%%%%%%%%%%%%%%%%%%%%
\begin{abstract}

We present ultravioliet (UV) observations of supernovae (SNe) obtained 
with the UltraViolet/Optical Telescope (UVOT) on board the {\sl Swift} spacecraft.  
This is the largest sample of UV light curves from any single instrument 
and covers all major SN types and most subtypes.
The UV light curves of SNe Ia are fairly 
homogenous while SNe Ib/c and IIP show more variety in their light curve shapes.  
The UV-optical colors clearly differentiate SNe Ia and IIP, particularly at early times.  
The color evolution of SNe IIP, however, makes their colors similar to SNe Ia 
at about 20 days after explosion.  SNe Ib/c are shown to have varied UV-optical colors.  
The use of UV colors to help type SNe will be important for high redshift SNe discovered 
in optical observations.
These data can be added to ground based optical 
and near infrared data to create bolometric light curves of individual objects and 
as checks on generic bolometric corrections used in the absence of UV data.  
This sample can also be 
compared with rest-frame UV observations of high redshift SNe observed 
at optical wavelengths.

\end{abstract}
%%%%%%%%%% Keywords %%%%%%%%%%%%%%%%%%%%%%%%%%%%%%%%%%%%%%%%%%%%%%%%%%%%%%%%%%%%
%%%%%%%%%%%%%%%%%%%%%%%%%%%%%%%%%%%%%%%%%%%%%%%%%%%%%%%%%%%%%%%%%%%%%%%%%%%%%%%%%
%\clearpage

%%%%%%%%%%%%%%%%%%%%%%%%%%%%%%%%%%%%%%%%%%%%%%%%%%%%%%%%%%%%%%%%%%%%%%%%%%%%%%%%%
\section{Ultraviolet SN Observations\label{IaOpt}}

From the earliest photon signal from a SN during the shock breakout, 
the UV light from supernovae contains many clues about the explosion 
and the environment, with application to both nearby and high-redshift SNe.
The contribution of UV light to the bolometric luminosity can be significant, 
particularly at the earliest epochs when the high temperature yields 
a large UV flux.  Because line blanketing in the UV is dominated 
by iron peak elements, the UV brightness is 
sensitive to the ionization level (cf. \citealp{Dessart_etal_2008}) 
and differences in metallicity \citep{Nugent_etal_1997,Lentz_etal_2000}.  
The large UV extinction observed in most extinction curves also allows 
UV observations to better constrain the reddening to individual objects 
\citep{Jeffery_etal_1994}.

%%%%%%%%%%%  UV SN Observatories

The first observation of UV light from a SN was performed by the 
Orbiting Astronomical Observatory in 1972, revealing the 
UV faintness of a type I SN \citep{Holm_etal_1974}.  
The International Ultraviolet Explorer (IUE) 
built up a larger sample with UV spectroscopic observations 
of 25 SNe \citep{CTF1995}, with excellent light curves for 
SNe IIL 1979C \citep{Panagia_etal_1980} and 1980K, the Ib 1983N, 
Ia 1992A \citep{Kirshner_etal_1993}, 
as well as the II-pec 1987A \citep{Pun_etal_1995}.  
The spectacular event of SN~1987A was also observed in the UV by the 
Astron Station \citep{Lyubimkov_1990}.  The Hubble Space Telescope (HST) 
has added excellent UV spectroscopic and photometric data for 
another $\sim$30 SNe (see \citealp{Panagia_2003} for a review of IUE and HST UV 
observations up to that time).
The Galaxy Evolution Explorer (Gal-Yam et al., in preparation), 
and the Optical Monitor on the XMM-Newton mission 
(cf. Immler et al. 2005) have added a few more observations each.  
Rest-frame UV observations of higher 
redshift SNe observed in the optical from the ground are also now being 
regularly obtained \citep{Astier_etal_2006, Foley_etal_2007, Ellis_etal_2008}.

The latest UV observatory is {\sl Swift} UVOT 
\citep{Roming_etal_2005}.  
The {\sl Swift} observatory's quick response capability and short term scheduling, 
necessitated by the unpredictable and variable behavior of individual 
gamma ray bursts (GRB), also allows newly discovered 
SNe to be observed quickly and with well sampled light curves \citep{Gehrels_etal_2004}.  
Data are available in a matter of hours, and observing
 times and filter combinations can be changed on a day to day basis 
depending on what is seen in recent observations.
Thus far, {\sl Swift} has focused on nearby SNe ($z\lesssim0.02$) for which 
high quality data can be obtained with only a small impact on spacecraft 
operations and {\sl Swift}'s primary mission to detect and observe GRBs.  
SN observations are performed under {\sl Swift}'s Target of Opportunity (ToO) program
\footnote{http://www.swift.psu.edu/too.html}.
A dedicated website\footnote{see http://swift.gsfc.nasa.gov/docs/swift/sne/swift\_sn.html}
 has been set up that gives the status of Swift SN observations, images and
regularly updated light curves 
for the benefit of the community.

In this paper we present some of the overall UV properties of SNe 
seen in our sample.  In particular, we contrast the UV light curve shapes 
and colors of the different types, which is best done with a sample observed 
with the same instrumental setup.  Understanding the brightness 
(relative to the optical) and temporal 
behavior should assist future observatories to understand the sampling and depth necessary to 
characterize a UV light curve.  

%%%%%%%%%%%%%%%%%%%%%%%%%%%%%%%%%%%%%%%%%%%%%%

\section{Observation Summary\label{sample}}

{\sl Swift} UVOT has observed 48 SNe 
between March 2005 \citep{Brown_etal_2005} and August 2007.  
UVOT typically observes SNe with three UV 
and three optical broadband filters.  The central wavelengths and 
widths are given in Table 1 \citep{Poole_etal_2007}.  
\citet{Brown_etal_2007b} 
display the transmission of these filters with respect to SNe spectra.
However, like IUE, they are limited to the brightest 
epochs of the the nearest SNe.  Here we focus on results from the 
photometry; spectroscopic results will be presented elsewhere 
(Bufano et al. in prep).

%%%%%%%%%%%%%%%%%%%%%%%%%%%%%%%%%%%%%%%%%%%%%%%%%%%%%%%%%%%%%%%%
\begin{deluxetable}{cccccc}

\tablecaption{{\sl Swift} UVOT Filter Characteristics}
\tablehead{\colhead{Filter} & \colhead{$\lambda_{central}$} & 
\colhead{FWHM} & \colhead{Zeropoint}  \\ 
\colhead{} & \colhead{(\AA)} & \colhead{(\AA)} &
\colhead{(mag)} }

%% All data must appear between the \startdata and \enddata commands
\startdata
uvw2 & 1928 & 657 & $17.35 \pm 0.03$\\
uvm2 & 2246 & 498 & $16.82 \pm 0.03$ \\
uvw1 & 2600 & 693 & $17.49 \pm 0.03$\\
$u$ & 3465 & 785 & $18.34 \pm 0.020$\\
$b$ & 4392 & 975 & $19.11 \pm 0.016$\\
$v$ & 5468 & 769 & $17.89 \pm 0.013$\\
\enddata

\end{deluxetable}

%%%%%%%%%%%%%%%%%%%%%%%%%%%%%%%%%%%%%%%%%%%%%%%%%%%%%%%%%%%%%%%%%%%%%%%%%%%%%%%%%

%%%%%%%%%%%%%%%%%%%%%%%%%%%%%%%%%%%%%%%%%%%%%%%%%%

Images were obtained from the {\sl Swift} archive, and for those whose most recent 
processing occurred prior to 2007, the raw images and event lists 
were reprocessed, primarily to utilize an improved plate scale for the uvw2 images 
and corrections to exposure times in the headers.  
Exposures were aspect corrected then coadded by epoch, usually corresponding 
to unique observation numbers.  Due to the rapid evolution of SN~2006aj/GRB060218 
during the first day the exposures in the first observation number were used 
individually and subsequent observations coadded as usual.
We performed aperture photometry using a 3\arcsec~source aperture 
to improve the signal to noise and minimize galaxy contamination compared 
to larger apertures (cf. \citealp{Li_etal_2006}).  A 
background region was chosen by eye to be similar to the galaxy background 
near the SN.    Over or undersubtraction of the underlying galaxy leads to the 
magnitudes being under or overestimated when the SN fades close the the galaxy level, 
so here we focus on the brightest epochs and less contaminated events where the 
galaxy contribution is less significant.  
Counts within a 5\arcsec~aperture around the SN were used 
to compute the coincidence loss correction factor (to be consistent with the calibration), 
and an average point spread function (psf) for 
each filter was used to compute the aperture correction factor from our 
3\arcsec~aperture to the 5\arcsec~for which the zeropoints are calibrated 
\citep{Poole_etal_2007}.  
The light curves of those with at least 3 UV detections well above the estimated 
galaxy background (26 out of the 48 SNe observed during this period) 
are displayed on a continuous timeline in Fig.\ 1, with individual light curves displayed in Figs. 2 and 3.  
 Fig.\ 1 illustrates the dynamic range and frequency of Swift SN observations.  
The bright limit, above which coincidence losses are hard to correct,  
is approximately 12th magnitude in the UV. In practice this saturation limit 
from a bright SN or a SN on a bright galaxy background has affected the optical light curves of 
some of these SNe but not the UV light curves.  
The faint limit can be as deep as about 21 mag 
but observations are typically terminated once the galaxy light dominates within 
the aperture.

%%%%%%%%%%%%%%%%%%%%%%%%%%%%%%%%%%%%%%%%%%%%%

%%%%%%%%%%%%%%%%%%%%%%%%%%%%%%%%%%%%%%%%%%%%%%%%%%

%%%%%%%%%%%%%%%%%%%%%%%%%%%%%%%%%%%%%%%%%%%%%%%%%%%%%%%%%%%%%%%%%%%%%%%%%%%%%%%%%

\section{Light Curves}

The temporal behavior of the UV light varies with SN type.  Fig.\ 2 shows UVOT 
lightcurves for the individual SNe Ia.  SNe Ia rise to a 
maximum in the UV peaking just before the optical. Similar to the optical, the UV brightness decays 
first steeply and then shallower due to radioactive decay of Nickel and Cobalt.  
The uvm2 photometry is much fainter than the other bands, so the points have larger errors, 
but the light curves seem to be 
broader (decaying shallower) than the other UV filters.
In Table 2 we report the typical decay rate in mags/100 days as in \citet{Pskovskii_1967}, 
with $\beta$ representing the early decay and $\gamma$ the later shallow decay.
The UV lightcurves of SNe Ia are fairly uniform, and the uvw1 curves match 
well with the HST/IUE spectrophotometry of SN~1992A in the comparable F275W band 
\citep{Kirshner_etal_1993,Brown_etal_2005,Milne_etal_2007}.  
More details on the lightcurves and generation 
of UV lightcurve templates for SNe Ia will be presented in Milne et al. (in preparation).

From a sample of light curves one can begin 
to discern what is normal and what is peculiar behavior.  Two SNe Ia that stand 
out are SNe 2005hk and 2005ke.  
SN~2005hk was already fading in the UV when {\sl Swift} observations began, nearly 10 
days before the optical maximum.  
SN~2005ke followed the typical Ia decay until about 15 days after maximum light 
when the UV brightness remained 
nearly constant for $\sim$20 days before fading again.  
In conjunction with a marginal x-ray 
detection, this plateau in the UV light curves has been attributed to interaction with the 
circumstellar material \citep{Immler_etal_2006}, though other causes such as 
reduced line blanketing have been suggested \citep{Kasliwal_etal_2008}.  
Contamination could 
be ruled out for SN~2005ke because the first observation showed the SN fainter 
than the plateau level and subsequent observations showed it had faded again. 
For other SNe contamination is harder to rule out, but SNe 2007ax \citep{Kasliwal_etal_2008} 
and 2006mr show hints of extended emission above the background.
SN~2006E also stands out with its relatively flat light curves, but that is because 
the peak and early decay were missed and the  
slow decay merely represents the shallower late time decay also seen in other SNe Ia.

For SNe Ib/c, the sample of well observed SNe is much smaller, with 3 light curves 
in our sample, but they appear to be as diverse in the UV as they are in the optical (see Fig.\ 3).
Other SNe Ib/c were also observed but for only a single epoch or were not well detected 
(see \citet{Holland_etal_2007} for additional SNe Ib/c observed during {\sl Swift}'s first two years).
It is hard to define a generic 
UV behavior, so instead we briefly describe each well sampled SN.

SN~2006aj, a Ic SN, was 
discovered following {\sl Swift} BAT trigger on GRB060218 \citep{Campana_etal_2006}.  
The UV initially rose rapidly reaching 
a bright peak about half a day after the trigger.  
This first peak has been attributed to the shock breakout from a dense 
stellar wind \citep{Campana_etal_2006,Blustin_2007,Waxman_etal_2007} or
self-absorbed synchrotron radiation \citep{Ghisellini_etal_2007}.  
This faded rapidly 
in the UV, plateauing briefly from 4$--$10 days after 
the trigger as the radioactive decay powered a supernova light curve, 
and then faded again.  In uvw2 and uvm2 this 
SN peaked 4 magnitudes fainter than the earlier peak.  
The optical curves are consistent with a SN Ic of intermediate luminosity 
between normal SNe Ic and previous GRB-associated, overluminous SNe Ic 
(cf. \citealp{Pian_etal_2006} and \citealp{Modjaz_etal_2006}).

SN~2006jc was a peculiarly bright and blue SN Ib \citep{Foley_etal_2007}, 
and the UVOT grism spectra 
show an indication of MgII emission \citep{Immler_etal_2008}.  
Discovered near maximum light, the UV and optical light curves all 
fade steeply and then shallower and then steeper again.  The UVOT light curves 
will be studied in more detail in Modjaz et al. (in preparation).
SN~2007Y, a peculiar Ib/c with spectral similarities to SN~2005bf \citep{Folatelli_etal_2007}, 
behaved more like a Ia in shape and color with the UV 
brightness rising with the optical, peaking a little sooner and fading 
slightly quicker. 

SNe IIP start off very bright and blue.  Brightening to a maximum just a 
few days after explosion, the UV magnitudes then fade rather linearly, 
with the uvm2 filter decaying slightly faster than the others.  Individual 
light curves of the SNe IIP are displayed in Fig.\ 3.  
This rapid drop in the UV brightness is driven 
by the cooling photosphere and the resulting line blanketing of heavy 
elements (cf. \citealp{Brown_etal_2007a}).   \citet{Dessart_etal_2008} compare the UVOT 
photometry of SNe II 2005cs and 2006bp with model spectra and demonstrate 
the usefulness of UV photometry in constraining the extinction and the 
temporal change in temperature and ionization.  
The optical brightness remains constant, resulting in a 
steady reddening with time. While this reddening is common 
to all the SNe IIP observed by UVOT, the decay slopes vary by nearly a 
factor of 2 (as shown in Table 2),
possibly reflecting the different cooling rates of the SN photospheres.

In order to better contrast the SN types, Fig.\ 4 shows the UVOT light curves 
of a well observed example of each type.  For easy comparison, the time and 
magnitude axis are the same across the plots and the magnitudes are unshifted 
to show the relative colors.  The explosion dates for SNe 2007af and 2007Y 
assume a rise time of 18 days to the maximum light in the V band 
for SNe Ia/b/c \citep{Garg_etal_2007, Stritzinger_etal_2002}, 
while SN~2006bp uses the explosion date determined by \citet{Dessart_etal_2008}.  
The main SN subtypes missing from our sample are SNe IIL, IIn, and IIb.  
Fortuitously, examples of these classes, along with the peculiar SN II 1987A 
were well observed with IUE and HST, making these sets complementary.  

%%%%%%%%%%%%%%%%%%%%%%%%%%%%%%%%%%%%%%%%%%%%%%
\begin{deluxetable}{ccccc}
\tablecaption{Decay Slopes}
\tablehead{\colhead{Filter} & \colhead{SN Ia $\beta$ } & 
\colhead{SN Ia $\gamma$ } & \colhead{SN Ibc $\beta$ } & \colhead{SN II  $\beta$ } \\ 
\colhead{} & \colhead{(mag/100 d)} & \colhead{(mag/100 d)} & 
\colhead{(mag/100 d)} & \colhead{(mag/100 d)}} 

\startdata
uvw2 &  9$--$10 &  3 &  18 &  23$--$37\\
uvm2 &  7$--$9 &  3 &  14 &  23$--$41\\
uvw1 &  11$--$13 &  2 &  18 &  20$--$32
\enddata

\tablecomments{
SNe used: Ia $\beta$ (SNe 2007af, 2005ke), SN Ia $\gamma$ (SN~2006E),  
SN Ibc  $\beta$  (SN~2007Y), SN II $\beta$ (SNe 2005cs, 2006at, 2006bp) }
\end{deluxetable}

%%%%%%%%%%%%%%%%%%%%%%%%%%%%%%%%%%%%%%%%%%%%%%

%%%%%%%%%%%%%%%%%%%%%%%%%%%%%%%%%%%%%%%%%%%%%%%%%%%%

\section{Colors}

In addition to physical properties like temperature and extinction, 
UV and UV-optical colors are also useful in differentiating SN types.  The use of optical 
colors to distinguish SNe of different types has been explored by 
multiple authors \citep{VandenBerk_etal_2001, Poznanski_etal_2002, 
Gal-Yam_etal_2004, Sullivan_etal_2006, Kuznetsova_Connolly_2007, 
Poznanski_Maoz_2007}.   These techniques are of primary importance 
for large and deep surveys and searches for which the SN candidates 
are either too faint or too numerous for spectroscopic classification 
of all candidates.  The upcoming Pan-STARRS could discover on the order of 
10,000 SNe/year\footnote{http://pan-starrs.ifa.hawaii.edu/project/reviews\\/PreCoDR/documents/scienceproposals/sne.pdf} and 
the Large Synoptic Survey Telescope 250,000 SNe/year.\footnote{http://www.lsst.org/Science/fs\_transient.shtml}  
Thus photometric measures 
of the SNe will be critical 
to classifying SNe, as well as determining photometric redshifts.  Rest frame 
UV observations can greatly improve the accuracy of such determinations.  
 Fig.\ 6 compares the color-color location of two SNe Ia (SNe 2007af and 2005ke, 
well observed examples of a normal and sub-luminous Ia respectively) and two SNe IIP 
(the well observed SNe 2005cs and 2006bp) in the 
5 colors (using neighboring filter combinations) available from the UVOT observations.  
Extinction vectors, corresponding to a color excess E(B-V)=0.1 and the 
Milky Way extinction law \citep{Cardelli_etal_1989} evaluated at the central wavelengths, 
are plotted in each color-color plot, though the effect of extinction can vary  
with different source spectra and choices of extinction laws.
As the UVOT $u,b,$ and $v$ magnitudes are similar to the Johnson $UBV$, the bottom 
right panel is comparable to Fig.\ 3 in \citet{Poznanski_etal_2002}.  In those colors 
there is still a lot of overlap between SNe Ia and II.  As one 
adds progressively bluer filters, SNe Ia and II are better separated, showing the 
advantage of rest-frame UV observations.  In the upper left panel, SNe Ia are actually 
bluer than SNe II in the uvw2-uvm2 color, but this is likely due to the red tail of the uvw2 
filter allowing more optical photons through and making the uvm2 look fainter by comparison.
As discussed more below, the UV and UV-optical colors of SNe II evolve dramatically, 
making their colors more similar to SNe Ia.

This clear difference between the UV-bright SNe II and the UV-faint SNe Ia  
was noticed after a few IUE observations of SNe Ia and II (\citealp{Panagia_1982}, 
see also \citealp{Panagia_2003}).  This "UV deficit" is caused by the very 
red Ia spectrum between $\sim2500$ and 4000 \AA~
(cf.\citealp{Kirshner_etal_1993}), and was exploited 
by \citet{Riess_etal_2004} to identify SNe Ia in the Hubble Deep Field.  
The left panel of Fig.\ 6 uses filters on either side and in the middle of that 
red spectral slope to give two colors, uvm2-uvw1 and uvw1-b, which best show the 
contrast between SNe Ia and II.  The SNe II 
are clearly separated into the top left portion of the plot, while the SNe Ia 
are in the bottom right.  The uvw1-b color is sufficient by itself to make the 
distinction, with uvw1$--$b$\la$1 corresponding to young SNe II.  
The right panel of Fig.\ 6 adds the colors of our 
three SNe Ib/c, spanning the range of colors seen in SNe Ia and II, 
complicating their differentiation.  The similarities of SN~2006jc with SNe IIn 
\citep{Pastorello_etal_2008} explains the blue 
UV-optical colors.  More typical SNe Ib/c have lower effective 
temperatures and the subsequent UV line blanketing make their colors more similar 
to SNe Ia. The contamination of SNe Ib/c from cosmological samples of SNe Ia, 
as discussed in \citet{Riess_etal_2004}, is reduced because 
the SNe Ib/c are less common and usually much fainter than SNe Ia. 

While red UV-optical colors differentiate well between young SNe II and Ia, 
they are not conclusive, as SNe II become redder with time.  
In Fig.\ 7 we display the uvw1-b color evolution of these same SNe.  
To determine the explosion date we have assumed a rise time of 18 days to 
the maximum light in the V band for SNe Ia \citep{Garg_etal_2007} 
and for the SNe II we use the explosion times 
determined by \citet{Dessart_etal_2008}.  At early times, there is a 
3 magnitude difference between the uvw1-B color of SNe Ia and II.  
However, this large difference does not persist as the SN II temperature 
(and correspondingly the UV flux) drops with time creating a redder spectrum 
whose colors become similar to SNe Ia (whose red colors evolve rather slowly) by day 20.
\citet{Fransson_etal_1987} noticed this rapid reddening in SN~1987A and 
cautioned against using the UV alone to distinguish SNe I from II.  Even a 
rough determination of the SN epoch, through either the light curve behavior 
or even the cadence of the SN search 
observations can help make the distinction.  Reddening is a further degeneracy 
which can also be broken 
by monitoring the color evolution, as reddening would cause a shift in 
the colors at a given epoch but not mask the differing slopes of the color evolution.

Understanding the UV color differences within and across types is 
especially important for classifying high redshift SNe.  Some UV information is 
being incorporated into such photoyping (cf. \citealp{Riess_etal_2004,Johnson_Crotts_2006}) 
but in the past has been limited due to the limited epochs at which UV information 
is available.
The addition of this UV photometry should help improve 
the understanding of the diversity and temporal change of the UV flux 
to allow SNe to be better identified at larger redshifts.  Other high redshift applications 
are discussed below.

%%%%%%%%%%%%%%%%%%%%%%%%%%%%%%%%%%%%%%%%%%%%%%%%%%%%%%%%%%%%
\section{High Redshift Application}

In addition to SN observations by UV satellites, deep optical observations will 
observe the rest-frame UV light from higher redshift SNe.  Fig.\ 8 displays 
the rest frame UV light sampled by the UVOT UV filters as it corresponds to 
observed wavelengths as a function 
of redshift, highlighting regions where the UV filters correspond well with commonly 
used optical and infrared filters.  To highlight the regions where the overlap 
between the filters is greatest, the bands are centered on the central wavelength 
and extend one quarter of the full width half max of the filter transmission in 
both directions \citep{Poole_etal_2007,Fukugita_etal_1996,Hewett_etal_2006}.  
This graphically depicts useful regions of 
overlap.  For example, the photons corresponding with the rest-frame uvw1 begins 
to be redshifted into the optical u' band 
(e.g. SDSS-II SN survey; \citealp{Sako_etal_2008}; and LSST) for objects at a redshift of $z\sim0.4$ 
and the g' band (e.g. SNLS; \citealp{Astier_etal_2006}) at $z\sim0.8$, where the 
u'-g' colors also correspond well with our uvw2$--$uvw1 colors.  While chasing the 
optical light to high redshifts will require observing into the infrared, 
the use of rest frame UV light can be done with optical and near-IR observations 
possible with current and planned large ground based telescopes.

\citet{Aldering_etal_2006} 
discuss many uses for rest-frame UV observations of these high redshift SNe 
in the context of SNAP, and this local sample should allow a comparison 
looking for evolutionary effects.  Since SNAP is planned to have 
logarithmically spaced filters beginning at 4000 \AA, they will cover the restframe 
UV observed by the UVOT filters for redshifts beyond z$\sim0.8$.  
More generally, any deep SN search 
optimized for finding and following high redshift SNe in the optical will likely 
also detect the rest-frame UV light of SNe (preferentially SNe II due to their 
brighter UV luminosities) at the high end of  
their target redshift and beyond.  
Measuring SN rates at higher redshifts is only one of many uses of these high redshift detections.  
Making full use of this data will require a better 
understanding of the UV light best obtained for nearby SNe for which multi-wavelength 
photometry and spectroscopy over a larger portion of the light curve is possible.
Rest-frame UV observations of a local sample of SNe have the further advantage 
of being a comparison sample with which to understand the high redshift SNe, 
look for evolutionary differences (see e.g. \citealp{Foley_etal_2007, Ellis_etal_2008}), 
and further constrain photometric redshifts, 
extinction, and luminosity distances.

%%%%%%%%%%%%%%%%%%%%%%%%%%%%%%%%%%%%%%%%%%%%%%%%%%%%%%

\section{Future Work}

We have presented here the largest collection of UV light curves obtained by 
any single instrument, which have allowed the study of individual objects as well 
as comparisons within and across SN types.  In addition to these apparent magnitudes, we are also 
working to calibrate the absolute magnitudes to many of these objects (Brown et al. 
in preparation).  
The absolute magnitudes can be used to study the utility of rest frame UV 
observations for cosmological measurements and for determining high redshift SN rates.  
We are also combining this data with optical and near-infrared observations to 
construct bolometric lightcurves that encompass more of the spectrum for individual objects 
 and refining bolometric corrections used in constructing bolometric curves 
from optical data alone (cf. \citealp{Contardo_etal_2000}).  
The light curve shapes and colors of our large sample should also help in the 
classification of SNe, particularly at higher redshifts when spectra are 
unobtainable and fewer rest-frame optical bands are observable. 
Future UV observatories, including a refurbished HST, TAUVEX \citep{Safonova_etal_2007}, 
WSO-UV \citep{Pagano_etal_2008} and others, as well as optical observatories observing 
the high redshift universe, can also benefit from these light curves 
to better understand and plan effective UV SN observations.

%%%%%%%%%%%%%%%%%%%%%%%%%%%%%%%%%%%%%%%%%%%%%%%%%%%%%%%%%%%%%%%%%%%%%%%%%%%%%%%%%

\acknowledgements
This work made use of public data from the {\sl Swift} data archive.
This work is supported at Penn State by NASA Contract NAS5-00136 
and {\sl Swift} Guest Investigator grant NNH06ZDA001N.
%%%%%%%%%%%%%%%%%%%%%%%%%%%%%%%%%%%%%%%%%%%%%%%%%%%%%%%%%%%%%%%%%%%%%%%%%%%%%%%%%

%%%%%%%%%%%%%%%%%%%%%%%%%%%%%%%%%%%%%%%%%%%%%%%%%%%%%%%%%%%%%%%%%%%%%%%%%%%%%%%%%

%%%%%%%%%%%%%%%%%%%%%%%%%%%%%%%%%%%%%%%%%%%%%%%%%%%%%%%%%%%%%%%%%%%%%%%%%%%%%%%%%

%%%%%%%%%%%%%%%%%%%%%%%%%%%%%%%%%%%%%%%%%%%%%%%%%%%%%%%%%%%%%%%%%%%%%%%%%%%%%%%%%
%\clearpage
%%%%%%%%%%%%%%%%%%%%%%%%%%%%%%%%%%%%%%%%%%%%%%%%%%%%%%%%%%%%%%%%%%%%%%%%%%%%%%%%%
%%%%%%%%%%%%%%  Master Light curve plot  %%%%%%%%%%%%%%%%%%%%%%%%%%%%%%%%%%%%%%%%

\clearpage
\begin{figure}
\epsscale{1.1}
\plotone{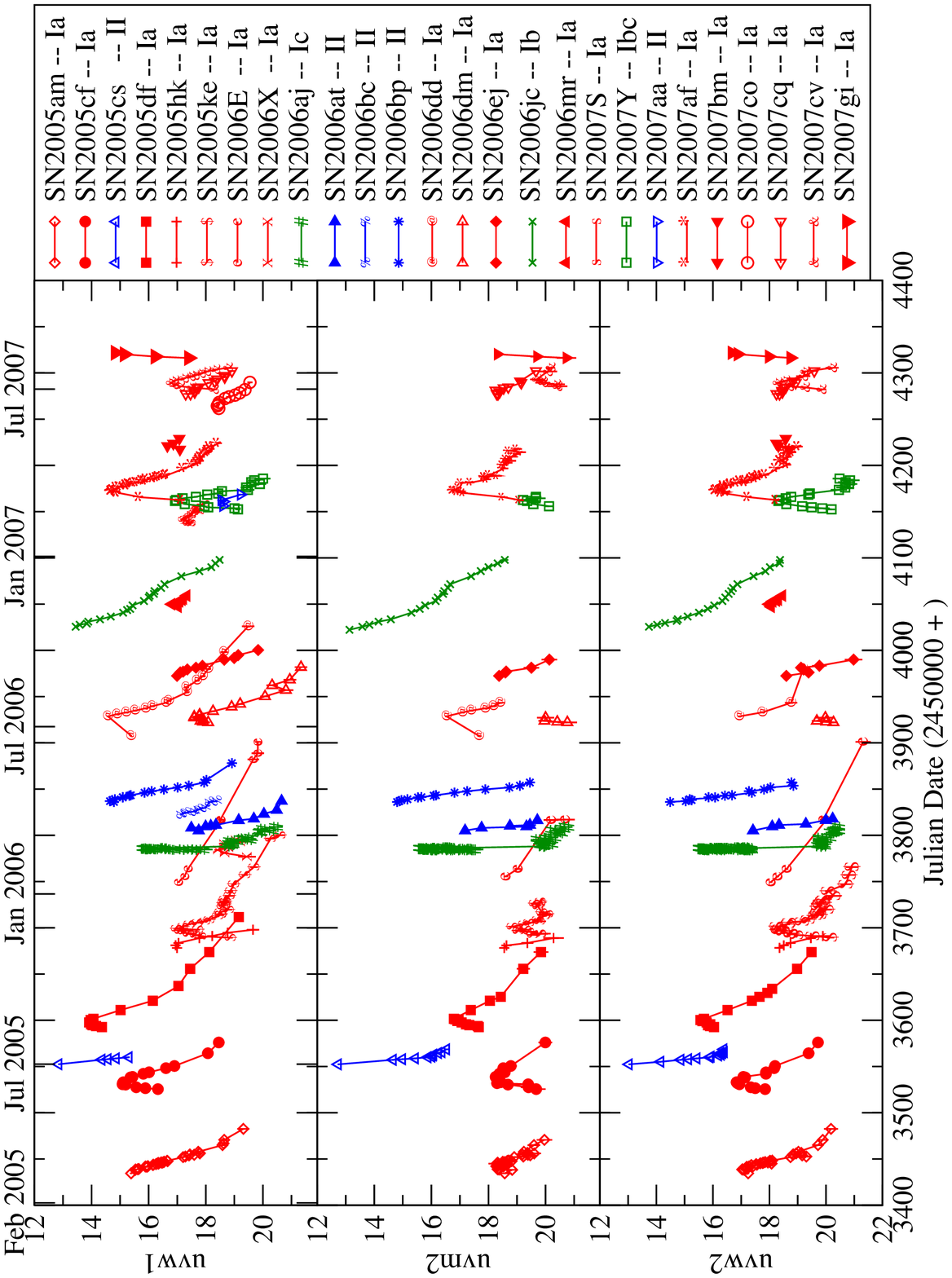}
\caption{UV light curves of SNe well detected in at least three epochs by UVOT.
SNe Ia are displayed in red, SNe Ib/c in green, and SNe II in blue.
    }\label{fig1}
\end{figure}
\clearpage
%%%%%%%%%%%%%%%%%%%%%%%%%%%%%%%%%%%%%%%%%%%%%%%%%%%%%%%%%%%%%%%%%%%%%%%%%%%%%%%%%

%%%%%%%%%%%%%%%%%%%%%%%%%%%%%%%%%%%%%%%%%%%%%%%%%%%%%%%%%%%%%%%%%%%%%%%%%%%%%%%%%
%%%%%%%%%%%%%% Individual Light curve plots          %%%%%%%%%%%%%%%%%%%%%%%%%%%%%%%%%%%%%%%

%\clearpage
\begin{figure}
\includegraphics[angle=0,scale=0.8]{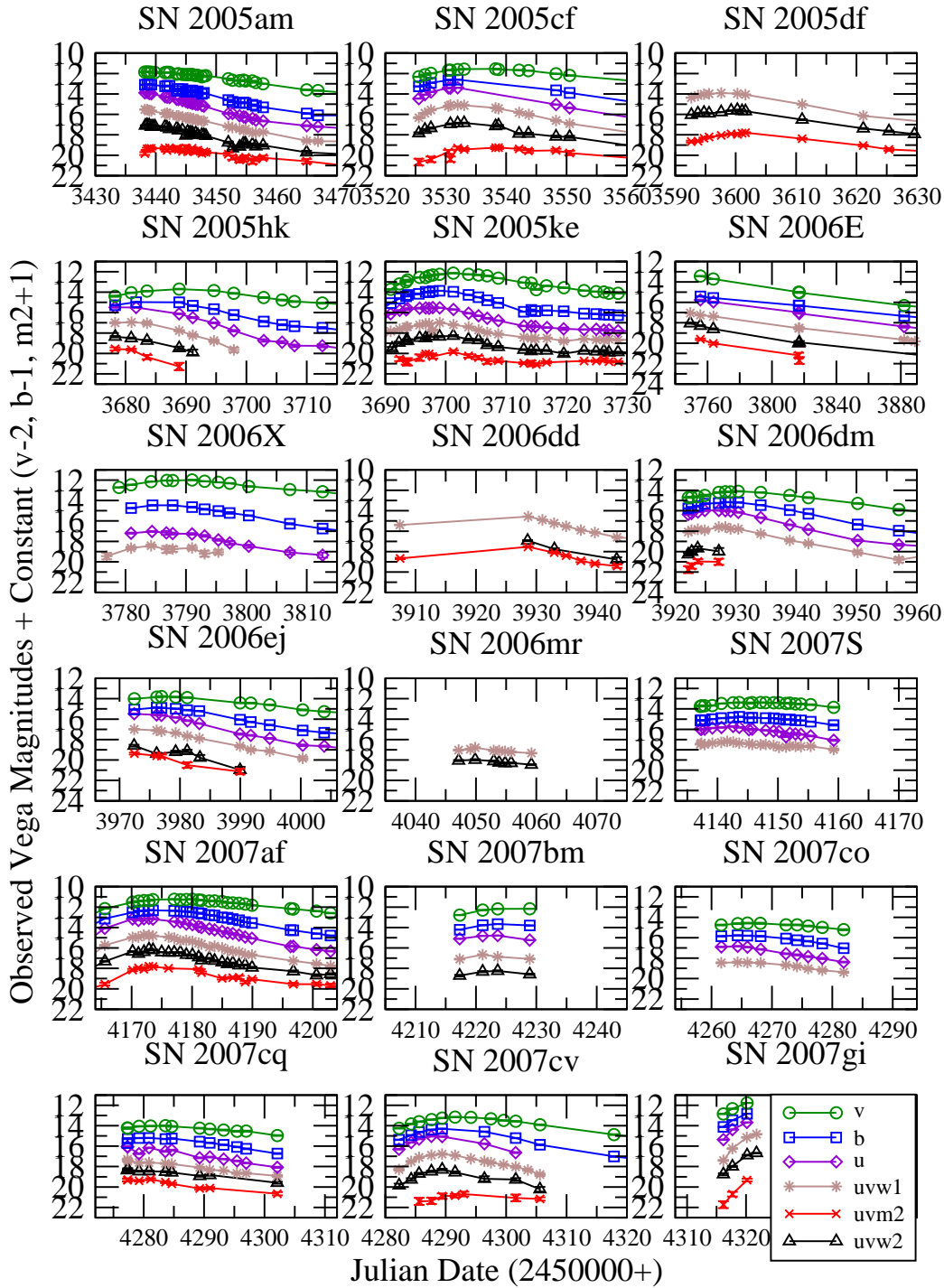}
\caption{UVOT light curves of 18 SNe Ia.  The curves have been offset by v-2, b-1, and uvm2+1 for clarity.  
For comparison purposes the axis are on the same scale, showing the 40 days around maximum light 
(with the exception of SN2006E which was only observed well after maximum).}\label{fig2}
\end{figure}
%\clearpage
%%%%%%%%%%%%%%%%%%%%%%%%%%%%%%%%%%%%%%%%%%%%%%%%%%%%%%%%%%%%%%%%%%%%%%%%%%%%%%%%%

%%%%%%%%%%%%%%%%%%%%%%%%%%%%%%%%%%%%%%%%%%%%%%%%%%%%%%%%%%%%%%%%%%%%%%%%%%%%%%%%%
%%%%%%%%%%%%%% Individual Light curve plots          %%%%%%%%%%%%%%%%%%%%%%%%%%%%%%%%%%%%%%%

%\clearpage
\begin{figure}
\epsscale{2.4}
%\rotate{270}
\includegraphics[angle=270,scale=0.7]{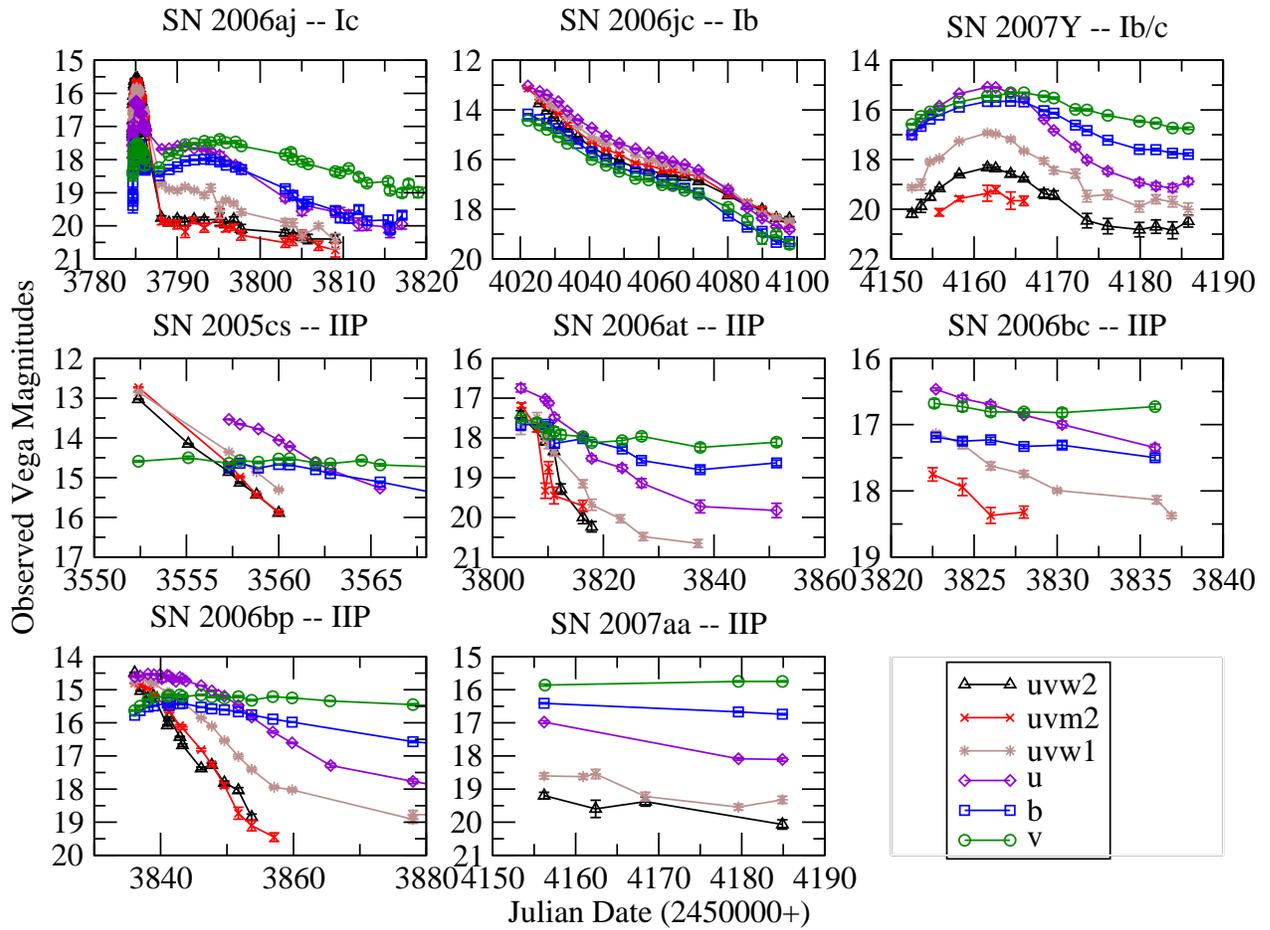}
%\plotone{SN2007af_6.ps}
\caption{UVOT light curves of 3 SNe Ib/c and 5 SNe II.}\label{fig3}
\end{figure}
%\clearpage
%%%%%%%%%%%%%%%%%%%%%%%%%%%%%%%%%%%%%%%%%%%%%%%%%%%%%%%%%%%%%%%%%%%%%%%%%%%%%%%%%

%%%%%%%%%%%%%%%%%%%%%%%%%%%%%%%%%%%%%%%%%%%%%%%%%%%%%%%%%%%%%%%%%%%%%%%%%%%%%%%%%
%%%%%%%%%%%%%% Type Comparison Light curve plots          %%%%%%%%%%%%%%%%%%%%%%%%%%%%%%%%%%%%%%%

%\clearpage
\begin{figure}
\epsscale{2.4}
\includegraphics[angle=270,scale=0.9]{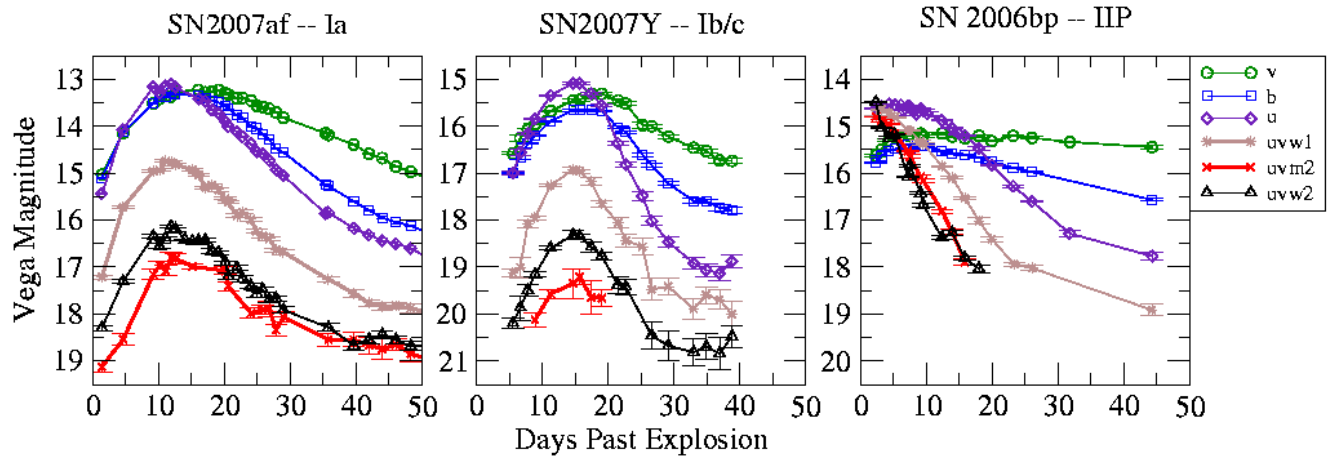}
\caption{UVOT light curves of SN~2007af (Ia), SN~2007Y (Ib/c), and SN~2006bp (IIP).  
The observed magnitudes have not been shifted and the horizontal and vertical axis 
are scaled the same to allow relative colors and slopes to be compared by eye.}\label{fig4}
\end{figure}
%\clearpage
%%%%%%%%%%%%%%%%%%%%%%%%%%%%%%%%%%%%%%%%%%%%%%%%%%%%%%%%%%%%%%%%%%%%%%%%%%%%%%%%%

%%%%%%%%%%%%%%%%%%%%%%%%%%%%%%%%%%%%%%%%%%%%%%%%%%%%%%%%%%%%%%%%%%%%%%%%%%%%%%%%%
%%%%%%%%%%%%%% Color color  plots          %%%%%%%%%%%%%%%%%%%%%%%%%%%%%%%%%%%%%%%

%\clearpage
\begin{figure}
%\epsscale{2.4}
%\rotate{270}
\includegraphics[angle=0,scale=0.8]{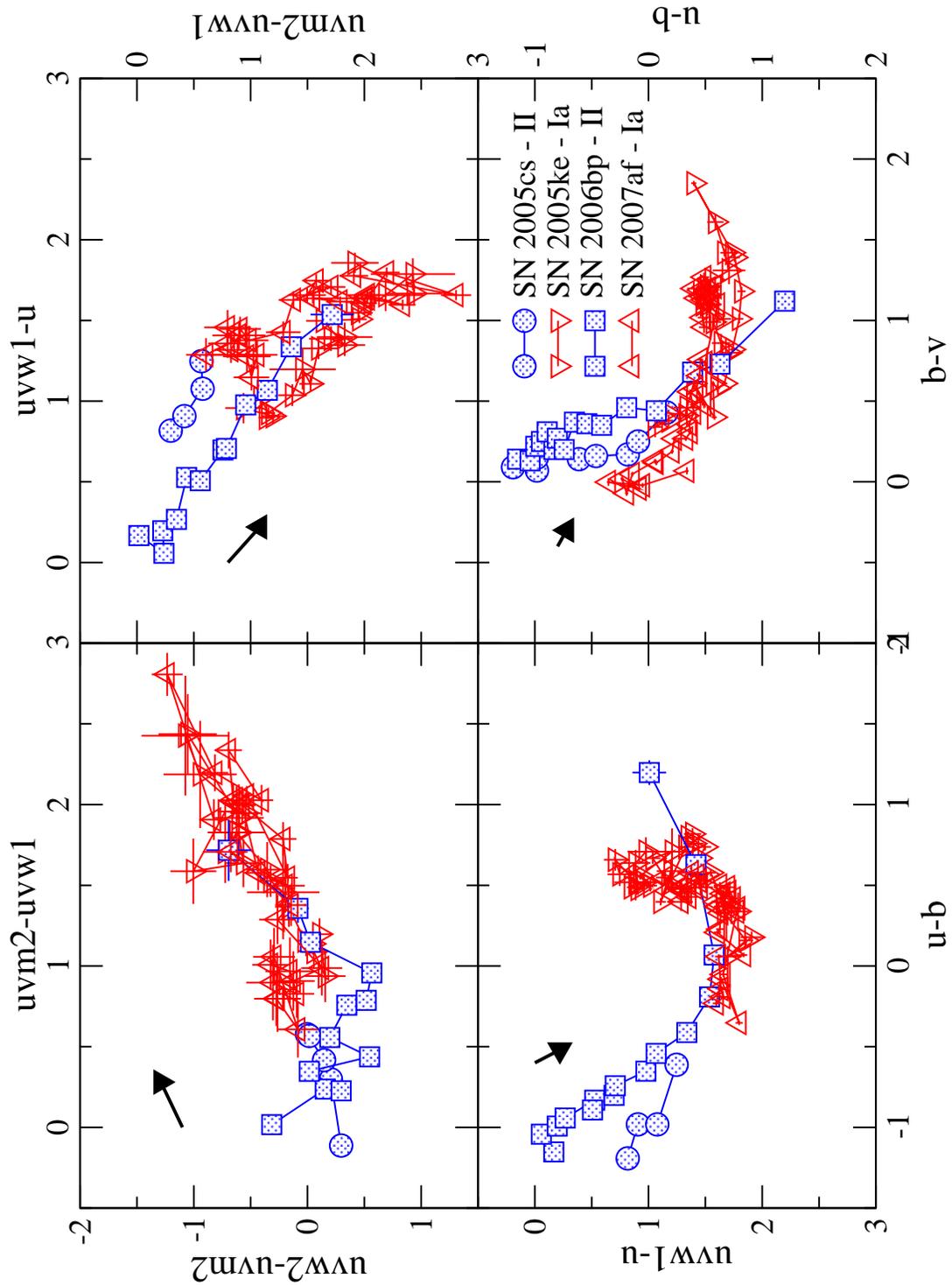}
\caption{Color-color plots showing the differentiation 
of young SNe II from SNe Ia.  The ability to differentiate the two increases 
at shorter wavelengths due to the contrasting spectral shapes.  The arrows 
are extinction vectors corresponding to E(B-V)=0.1 computed using the 
MW extinction relations of \citet{Cardelli_etal_1989} evaluated at the 
central wavelength of the UVOT filters.  Note that MW extinction actually causes 
bluer uvw2-uvm2 colors due to the uvm2 filter coinciding with the 2200 \AA bump in the 
MW extinction curve rather than the typical "reddening" effect of extinction.}\label{fig5}
\end{figure}
%\clearpage
%%%%%%%%%%%%%%%%%%%%%%%%%%%%%%%%%%%%%%%%%%%%%%%%%%%%%%%%%%%%%%%%%%%%%%%%%%%%%%%%%

%%%%%%%%%%%%%%%%%%%%%%%%%%%%%%%%%%%%%%%%%%%%%%%%%%%%%%%%%%%%%%%%%%%%%%%%%%%%%%%%%
%%%%%%%%%%%%%% uvm2-uvw1, uvw1-b  Color color  plots          %%%%%%%%%%%%%%%%%%%%%%%%%%%%%%%%%%%%%%%

%\clearpage
\begin{figure}
\epsscale{1.1}
%\rotate{270}
\plottwo{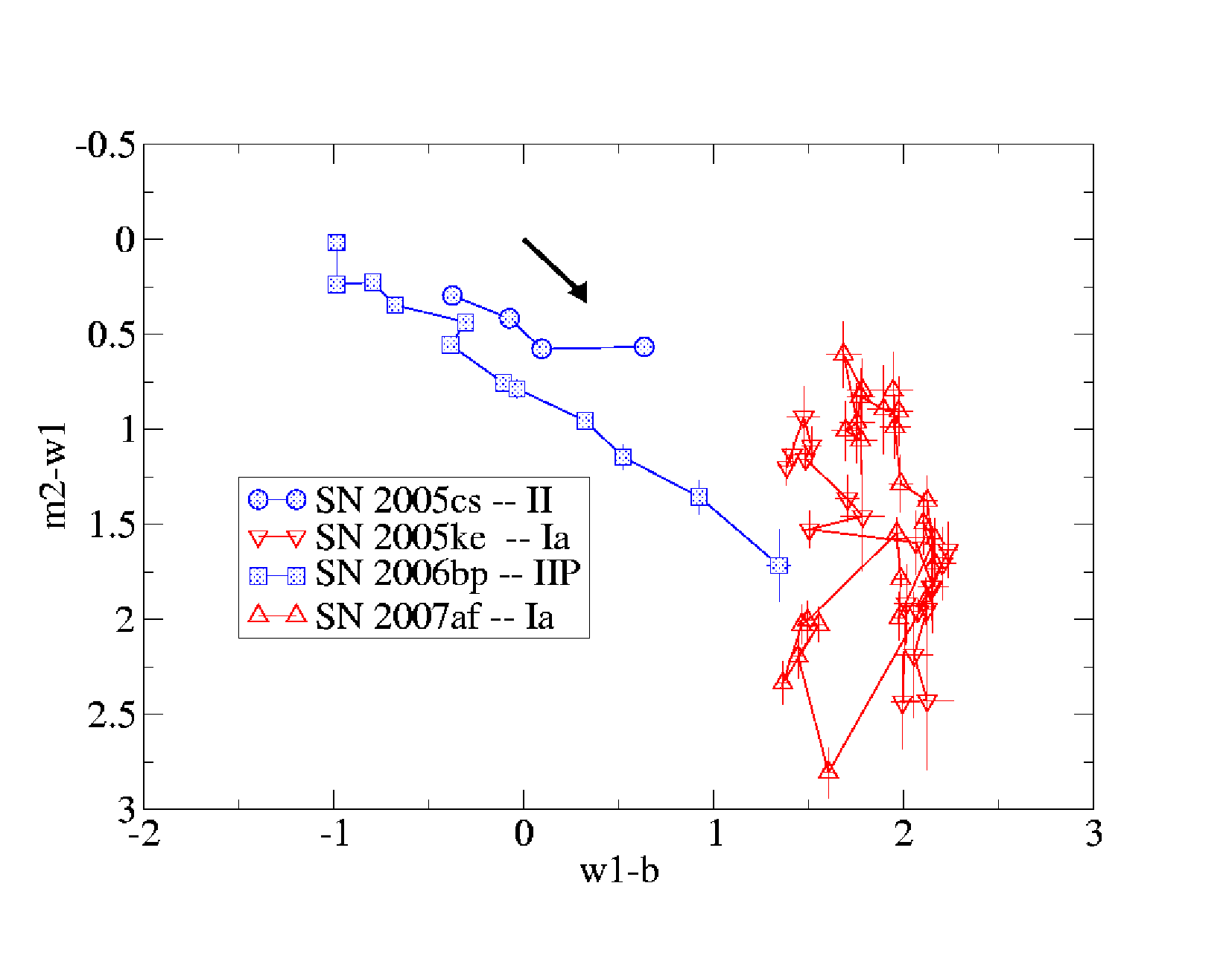}{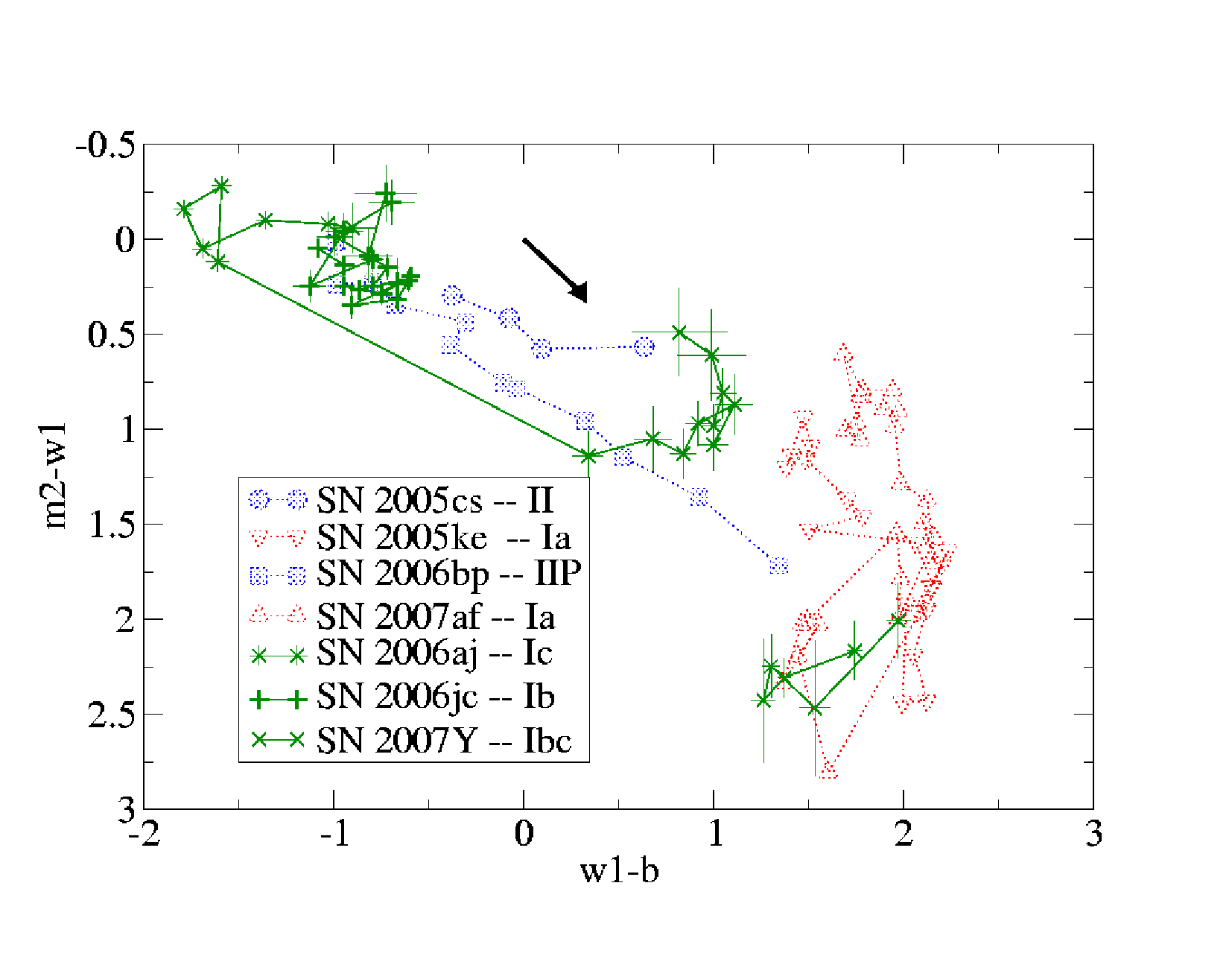} 
\caption{Left panel: uvm2-uvw1 v. uvw1-b color color plots showing the clear 
differentiation of young SNe II from SNe Ia. The SNe Ia are in red, 
 and the SNe IIP in blue.  The right panel shows the location 
of our 3 SNe Ib/c (marked in green) in the same color-color space.  
The arrows correspond to the uvw1-b color change of MW extinction corresponding 
to E(B-V)=0.1.  }\label{fig6}
\end{figure}
%\clearpage
%%%%%%%%%%%%%%%%%%%%%%%%%%%%%%%%%%%%%%%%%%%%%%%%%%%%%%%%%%%%%%%%%%%%%%%%%%%%%%%%%

%%%%%%%%%%%%%%%%%%%%%%%%%%%%%%%%%%%%%%%%%%%%%%%%%%%%%%%%%%%%%%%%%%%%%%%%%%%%%%%%%
%%%%%%%%%%%%%% Color evolution plots          %%%%%%%%%%%%%%%%%%%%%%%%%%%%%%%%%%%%%%%

%\clearpage
\begin{figure}
\includegraphics[angle=270,scale=0.8]{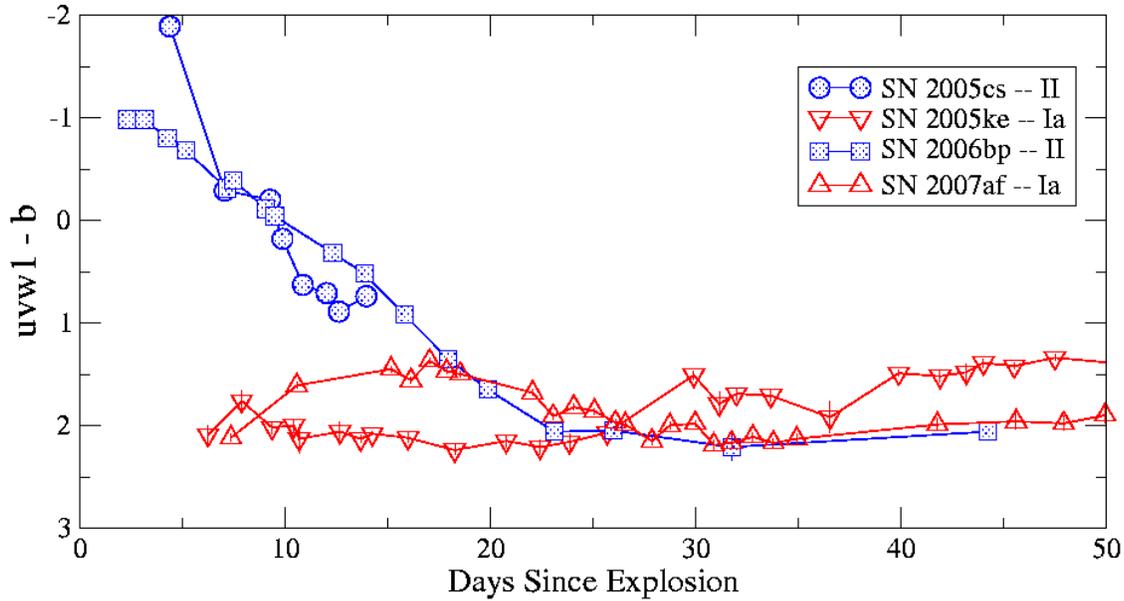}
\caption{UV-optical color evolution of two SNe Ia and two SNe IIP. The symbols are the same as used in the 
above color-color plots.  The dramatic color difference at early times vanishes by about 20 days after the explosion.}\label{fig7}
\end{figure}
%\clearpage
%%%%%%%%%%%%%%%%%%%%%%%%%%%%%%%%%%%%%%%%%%%%%%%%%%%%%%%%%%%%%%%%%%%%%%%%%%%%%%%%%

%%%%%%%%%%%%%%%%%%%%%%%%%%%%%%%%%%%%%%%%%%%%%%%%%%%%%%%%%%%%%%%%%%%%%%%%%%%%%%%%%
%%%%%%%%%%%%%% Filter -redshift plots          %%%%%%%%%%%%%%%%%%%%%%%%%%%%%%%%%%%%%%%

%\clearpage
\begin{figure}
\epsscale{2.4}
%\rotate{270}
\includegraphics[angle=0,scale=0.5]{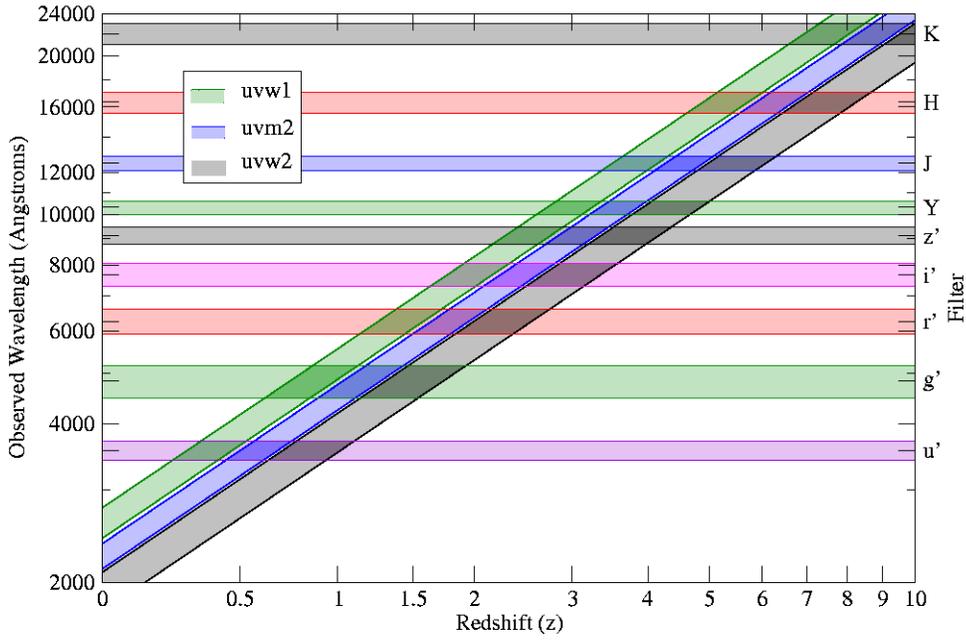}
\caption{Rest frame UV light sampled by the UVOT UV filters as it corresponds to 
observed wavelengths as a function of redshift.
The observed wavelength ranges for commonly used filter sets (optical--SDSS \citep{Fukugita_etal_1996}
and infrared--UKIRT set \citep{Hewett_etal_2006}) are also highlighted.
}\label{fig8}
\end{figure}
%\clearpage
%%%%%%%%%%%%%%%%%%%%%%%%%%%%%%%%%%%%%%%%%%%%%%%%%%%%%%%%%%%%%%%%%%%%%%%%%%%%%%%%%

\end{document}